\documentclass[runningheads]{llncs}
\usepackage{graphicx}
\usepackage{cite}
\usepackage{booktabs}
\usepackage{subcaption}
\usepackage{xcolor}
\usepackage{float}
\usepackage{siunitx}
\usepackage{amsmath,amsfonts,amssymb}
\usepackage{bm}
\usepackage{textgreek}

\graphicspath{{./figures/}}

\begin{document}	
\newcommand{\chist}{\chi_{\mathrm{st}}}
\newcommand{\Zst}{Z_{\mathrm{st}}}
\newcommand{\filt}[1]{\widetilde{#1}}
\newcommand{\Zfilt}{\filt{Z}}
\newcommand{\Zfvar}{\filt{Z''^2}}
\newcommand{\Cfilt}{\filt{C}}
\newcommand{\Cfvar}{\filt{C''^2}}
\newcommand{\rhofilt}{\overline{\rho}}
\newcommand{\Rey}{\mathrm{Re}}

\newcommand{\tsaf}[1]{\langle \filt{#1} \rangle}
\newcommand{\rmsf}[1]{\sqrt{\langle(\filt{#1}-\tsaf{#1})^2 \rangle}}
\newcommand{\means}[1]{{#1}_{\mathrm{mean}}}
\newcommand{\rmss}[1]{{#1}_{\mathrm{RMSD}}}

\newcommand{\di}{\,\mathrm{d}}
\newcommand{\lb}{\left(}
\newcommand{\rb}{\right)}
\newcommand{\ls}{\left[}
\newcommand{\rs}{\right]}

\newcommand{\nl}{ \notag \\ && &}
\newcommand{\nlp}{ \notag \\ && & \quad}
\newcommand{\nle}{ \\ \notag \\ &&}
\newcommand{\nlc}{ + \\ & \quad +}

\newcommand{\reff}[1]{Figure~\ref{#1}}
\newcommand{\refe}[1]{Equation~\eqref{#1}}
\newcommand{\refs}[1]{Section~\ref{#1}}
\newcommand{\refss}[1]{Subsection~\ref{#1}}
\newcommand{\reft}[1]{Table~\ref{#1}}
\newcommand{\refa}[1]{Appendix~\ref{#1}}
\newcommand{\rms}{RMSD }

\def\dd#1#2{\frac{\mbox{d} #1}{\mbox{d} #2}}
\def\pp#1#2{\frac{\partial #1}{\partial #2}}
\newcommand{\vect}[1]{\mathbf{#1}}
\newcommand{\vecg}[1]{\mathrm{\textbf{#1}}}
\newcommand{\avg}[1]{\langle{#1}\rangle}
\newcommand{\flu}[1]{{#1}'}
\newcommand{\mat}[1]{\mathbf{#1}}
\newcommand{\tens}[1]{\mathbf{#1}}

\newcommand{\abbrlabel}[1]{\makebox[3cm][l]{\textbf{#1}\ \dotfill}}
\newenvironment{abbreviations}{\begin{list}{}{\renewcommand{\makelabel}{\abbrlabel}}}{\end{list}}

\newcommand{\rebuttal}[1]{{#1}}
\newcommand{\press}[1]{{#1}}
\newcommand{\apress}[1]{{\color{blue}{#1}}}
\newcommand{\picsize}{\fontsize{10}{12}\selectfont}
\newcommand{\picbox}[1]{{#1}}

%
\title{Deep learning at scale for subgrid modeling in turbulent flows: regression and reconstruction}
%
\titlerunning{Deep learning at scale for subgrid modeling in turbulent flows}
%
%
%
\author{Mathis Bode\inst{1}\orcidID{0000-0001-9922-9742} \and
Michael Gauding\inst{2}\orcidID{0000-0003-0038-5249} \and
Konstantin Kleinheinz\inst{1}\orcidID{0000-0003-0569-2221} \and
Heinz Pitsch\inst{1}\orcidID{0000-0001-5656-0961}}
\authorrunning{M. Bode et al.}
%
\institute{Institute for Combustion Technology, RWTH Aachen University, Templergraben 64, 52062 Aachen, Germany \\
\email{\{m.bode,k.kleinheinz,h.pitsch\}@itv.rwth-aachen.de} \and
CORIA -- CNRS UMR 6614, Saint Etienne du Rouvray, France\\
\email{michael.gauding@coria.fr}}

\maketitle              

\begin{abstract}
	Modeling of turbulent flows is still challenging. One way to deal with the large scale separation due to turbulence is to simulate only the large scales and model the unresolved contributions as done in large-eddy simulation (LES). This paper focuses on two deep learning (DL) strategies, regression and reconstruction, which are data-driven and promising alternatives to classical modeling concepts. Using three-dimensional (3-D) forced turbulence direct numerical simulation (DNS) data, subgrid models are evaluated, which predict the unresolved part of quantities based on the resolved solution. For regression, it is shown that feedforward artificial neural networks (ANNs) are able to predict the fully-resolved scalar dissipation rate using filtered input data. It was found that a combination of a large-scale quantity, such as the filtered passive scalar itself, and a small-scale quantity, such as the filtered energy dissipation rate, gives the best agreement with the actual DNS data. Furthermore, a DL network motivated by enhanced super-resolution generative adversarial networks (ESRGANs) was used to reconstruct fully-resolved 3-D velocity fields from filtered velocity fields. The energy spectrum shows very good agreement. As size of scientific data is often in the order of terabytes  or more, DL needs to be combined with high performance computing (HPC). Necessary code improvements for HPC-DL are discussed with respect to the supercomputer JURECA. After optimizing the training code, \SI{396.2}{TFLOPS} were achieved.

	\bigskip
	
	\bigskip
	
\end{abstract}


\section{Introduction}
The turbulent motion of fluid flows poses some of the most difficult and fundamental problems in classical physics as it is a complex, strongly non-linear, multi-scale phenomenon~\cite{ruelle1971nature}. A general challenge in turbulence research is to predict the statistics of fluctuating
velocity and scalar fields and develop models for a precise
statistical prediction of these fields even in scale-resolved
simulations~\cite{piomelli1999large,kerstein2002turbulence}.
%

Large-eddy simulation (LES) is known to be a suitable modeling approach for turbulent flows and solves for the larger, flow-dependent
scales of the flow by modeling all scales below a particular filter
width~\cite{germano1991dynamic,leonard1975energy}. It is assumed that
the smaller, unresolved scales reveal certain universal features and
decouple from the larger non-universal scales. As a consequence,
models for LES can be built from relatively simple, semi-empirical
algebraic relations that are oftentimes based solely on dimensional
arguments~\cite{smagorinsky1963general}.  One approach to develop and
test such models is to perform fully resolved direct numerical
simulations (DNSs), filter the resulting data with a given filter
kernel, and find functional relations between the DNS results and the
filtered data. The objective of the present work is to move beyond
simple algebraic models for LES and use a data-driven approach with deep learning (DL) for modeling and reconstructing subfilter statistics for turbulent flows.

DL has gained immense interests from various industries and research groups in the age of big data. Prominent applications of DL include image processing~\cite{dong2014learning,wang2019,greenspan2016guest,wang2013learning}, voice recognition~\cite{hinton2012deep}, or  website customization~\cite{langheinrich1999unintrusive}. Reasons for that are the continued growth of computational power (especially GPUs) and the availability of exceptionally large
labeled experimental data sets. Also in the field of fluid mechanics and especially turbulence research, data-driven methods and DL have become more popular over the last years. However, often the applications are limited by either using only simple networks or small, artificial datasets.

Parish and Duraisamy~\cite{parish2016paradigm} used an approach called field inversion and machine learning (FIML), which moves beyond parameter calibration and uses data to directly infer information about the functional form of model discrepancies. They applied their approach to turbulent channel flows. Srinivasan et
al.~\cite{srinivasan2019predictions} assessed the capabilities of
neural networks to predict temporally evolving turbulent flows and
concluded that long short-term memory (LSTM) networks perform better
than multi-layer perceptron (MLP) approaches. Ling et al.~\cite{ling2016reynolds}
also presented a method using deep neural networks to learn a model
for the Reynolds stress anisotropy tensor from high-fidelity
simulations and experimental data. The Reynolds stress anisotropy
predictions were found to be more accurate than conventional Reynolds-averaged Navier-Stokes (RANS)
models, however the network could not perfectly reproduce the DNS
results.  Milano and Koumoutsakos~\cite{milano2002neural} modeled the near-wall region of turbulent
flows. Lapeyre et al.~\cite{lapeyre2019training} and Beck et
al.~\cite{beck2018neural} have documented the possibility of using
ML in designing subgrid-scale models for
LES. Maulik and
San~\cite{maulik2017neural} presented their use of a single-layer feedforward
artificial neural network (ANN) architecture trained through a supervised
learning approach for the deconvolution of flow variables from their
coarse-grained computations such as those encountered in LES. The subfilter-scale content recovery was
benchmarked against several popular structural closure modeling
strategies. Bode et al.~\cite{bode2018} studied the accuracy of various network architectures for predicting statistics of turbulent flows.
Machine learning (ML) and DL have also been applied to flow control~\cite{lee1997application,gautier2015closed}, development of low-dimensional models~\cite{shimizu2018construction}, generation of inflow conditions~\cite{fukami2019synthetic}, or structure identification in two-dimensional (2-D) decaying turbulence~\cite{jimenez2018machine}. Kutz~\cite{kutz2017} summarized more applications of DL in the field of fluid dynamics.

This work focuses on two different approaches in the context of data-driven turbulence modeling with DL: regression and reconstruction. In the regression part, a supervised learning method is used to
predict closure terms in the context of LES modeling based on filtered quantities. Simple ANNs are employed to predict, for example, the turbulent viscosity or the scalar dissipation rate. In the reconstruction part, a generative adversarial network (GAN) approach is followed to reconstruct fully-resolved turbulence
fields from filtered data. Results with respect to different network architectures and different quantities are discussed here. Furthermore, DL based on 3-D scientific data differs from DL on images not only in terms of the size of total data but also in the size of a single realization used for training. The size of scientific data can easily be in the order of hundreds of terabytes while training is traditionally performed with much smaller data. Therefore, DL on scientific data is often not possible without the usage of supercomputers and corresponding high performance computing (HPC) approaches. These computing aspects are also discussed in this work.

The remainder of this article is organized as follows. \refs{drr:sec:dsd} describes the used datasets. In \refs{drr:sec:M}, details about the regression and reconstruction methodologies are given, and results are discussed. Challenges with respect to computational aspects are addressed in \refs{drr:sec:C}. The paper finishes with conclusions.

\section{Dataset description}
\label{drr:sec:dsd}
The training and reconstruction is based on data obtained from
high-fidelity homogeneous isotropic forced turbulence simulations~\cite{gauding2017high,gauding2019self} in this work. The data was
generated by DNSs of the incompressible Navier-Stokes equations (NSEs) in a
triply periodic cube with size $2 \pi$ and $256^3$ collocation
points. Moreover, advection-diffusion equations of passive scalars were solved, which were used for tracking species or mixture factions.
Turbulence was kept in a statistically steady state by a large-scale
stochastic forcing scheme~\cite{eswaran1988examination}\rebuttal{, whereas the passive
scalars were forced by an imposed uniform mean gradient. The governing
equations were solved by an accurate pseudo-spectral approach with
integrating factor technique.}  A
pseudo-spectral approach with integrating factor technique was used
for accuracy. For efficiency, the non-linear transport term of the
NSEs was computed in physical space, and a truncation
technique with a smooth spectral filter was applied to reduce aliasing
errors. The library P3DFFT was used for the spatial decomposition and
to perform the fast Fourier transform. The code employs a hybrid
MPI/OpenMP parallelization and reveals a nearly linear scaling up to
two million threads.

Turbulence \rebuttal{in simple incompressible flows} can be characterized by a single characteristic number,
the Reynolds number $\Rey$, for example defined based on the Taylor length scale $\lambda$ as
\begin{equation}
  \Rey_\lambda = \frac{u' \lambda}{\nu},
\end{equation}
where $\nu$ is the kinematic viscosity and $u'$ is the root-mean-square deviation of the velocity vector $\vect{u}$. $u'$ is defined as
\begin{equation}
  \press{u' = \sqrt{\big<\frac{(\vect{u}-\big< \vect{u} \big>)\cdot(\vect{u}-\big< \vect{u} \big>)}{3}\big>}}
\end{equation}
with bold indicating tensors including vectors. Ensemble-averages are denoted by
angular brackets and computed over the full computational domain due to
statistical homogeneity of the DNS setup. \rebuttal{All velocity component fields are shifted to zero mean in this work as typically done for homogeneous isotropic turbulence.} The
Taylor-based Reynolds numbers of the used DNSs equal\rebuttal{s} approximately \num{43}, which
is large enough to ensure a non-linear transfer of turbulent energy
from the large, energy\rebuttal{-}containing scales toward the small, dissipative
scales.

The coarse-grained data was generated by applying a filter-kernel
$G(\vect{r})$ to the DNS data, i.\,e.
\begin{equation}
  \label{drr:eq:filter}
  \rebuttal{\bar{ \{\cdot\}}(\vect{x}) = \iiint \{\cdot\}(\vect{r}) G(\vect{x}- \vect{r})\di \vect{r}},
\end{equation}
where an overbar denotes filtered quantities.  For efficiency, the
filtering procedure is applied in spectral space, where a rotationally symmetric Gaussian
filter kernel, defined as
\begin{equation}
  \rebuttal{\hat{G}(\kappa) = \exp\left( - \frac{\kappa^2 \Delta^2}{24}  \right)}
\end{equation}
with $\kappa$ as the magnitude of the wavenumber vector $\boldsymbol{\kappa}$, is used. The Gaussian filter kernel is local in both spectral and real
space and avoids erroneous fluctuations in the filtered fields. The
cut-off wavenumber $\kappa_c$ is related to the filter-width $\Delta$
by
\begin{equation}
  \kappa_c = \frac{\pi}{\Delta}.
\end{equation}

In this paper, two statistically independent flow
\rebuttal{time-steps} (denoted by case A and case B) \rebuttal{with about two integral times in between} are studied. The filter width was
chosen as $\kappa_c=16$, which corresponds to a length scale at the
end of the restricted scaling range. Characteristic quantities of the
DNSs and the filtered data are given in \reft{drr:tab:dns}. Here,
$\big< k \big>$ denotes the ensemble-averaged turbulent \rebuttal{kinetic}
energy, $\big< \varepsilon \big>$ the ensemble-averaged dissipation rate of turbulent \rebuttal{kinetic} energy, and $\big< \chi \big>$  the ensemble-averaged dissipation rate of scalar \rebuttal{variance}. All quantities in this work are arbitrarily normalized without loss of generality.
\begin{table}
  \centering
  \begin{tabular}{ccc}
    \hline
  & Case A & Case B \\ \hline
  $\big< k \big>$ & 9.67 & 10.93\\
  $\big< \bar k \big>$ & 9.19 &  10.36\\
  $\big< \varepsilon \big>$ & 10.69 & 13.06\\
  $\big< \bar \varepsilon \big>$ & 8.62 & 10.11\\
  $\big< \chi \big>$ & 3.30 & 6.59\\
  $\big< \bar \chi \big>$ &2.16 & 4.11\\
  $\mathit{Re}_\lambda$ & 42.7 & 43.7\\ \hline
\end{tabular}
\caption{Characteristic properties of the DNSs and the filtered
  velocity and scalar field.}
  \label{drr:tab:dns}
\end{table}
\section{Modeling}
\label{drr:sec:M}
This section describes the regression and reconstruction approaches by showing results for two network architectures.  All networks were implemented using the Keras API~\cite{keras2019} built on the TensorFlow~\cite{abadi2015} backend.

\subsection{Regression}
The filtered NSEs contain unclosed terms which need to be modeled~\cite{bode2019bspline}. An often used closure for the filtered momentum equation relies on the \rebuttal{eddy-}viscosity $\nu_\mathrm{T}$ modeled as
\begin{equation}
  \label{drr:eq:nut}
\nu_\mathrm{T} = (C_\mathrm{s}\Delta)^2 \sqrt{\bar{\tens{S}}:\bar{\tens{S}}},
\end{equation}
where $C_\mathrm{s}$ is a model constant, $\Delta$ is the filter
width, and $\bar{\tens{S}}$ is the filtered rate of strain tensor defined as
\begin{equation}
  \bar{\tens{S}} =
  \frac{1}{2} \left(
    \nabla \bar{\vect{u}} + (\nabla \bar{\vect{u}})^\intercal \right)
    \label{drr:eq:S}
\end{equation}
with $\nabla$ being the del operator. Furthermore, the prediction of turbulent mixing requires an accurate prediction of the
\press{mean scalar dissipation rate $\big<\chi\big>$, which} is the sink term
in the transport equation of the \rebuttal{mean scalar variance $\big< (\phi-\big< \phi \big>)^2\big>$}\press{. Here, the local instantaneous scalar dissipation rate is} defined as
\begin{equation}
  \rebuttal{\chi = 2D \nabla(\phi-\big< \phi \big>)\cdot\nabla(\phi-\big< \phi \big>)},
    \label{drr:eq:sd_1}
\end{equation}
where $\phi$ denotes \rebuttal{the} transported scalar quantity, and $D$ is the molecular
diffusivity. \rebuttal{All scalars were shifted to zero mean in this work.} The mean scalar dissipation rate is related to the scalar
variance spectrum $E_\phi$ by
\begin{equation}
  \label{drr:eq:sd}
  \rebuttal{\big< \chi \big> = 2D \int_0^\infty \kappa^2 E_\phi(\kappa) \di \kappa,}
\end{equation}
\press{which} signifies that mainly the smaller
scales contribute to the mean scalar dissipation rate $\big< \chi
\big>$. As these scales are not available in coarse-grained fields or
LES, an accurate modeling of $\chi$ is \rebuttal{necessary}.

In the context of LES modeling, regression evaluated with neural network architectures can be used to obtain optimal predictions of subgrid quantities or contributions based on the incomplete information resolved in the LES. One example is to train a DL network with filtered DNS quantities as input and the corresponding DNS quantities as 'label' to learn the relation between the quantities resolved in LESs and their subgrid contributions. In the following subsections, this will be shown with simple feedforward ANNs. Unlike classical linear or logistic regression models, regression through neural
networks can represent more complex functions by data
manipulations through dense layers. The number of layers and the
number of nodes in each layer can be varied to obtain optimal networks
and results~\cite{ling2016reynolds}. Activation functions
in each layer can be used to add non-linearity to the
regression model, and a dropout layer can be added for regularization, so that
certain nodes are ignored during training to reduce overfitting or
high variance. In the next subsection, regression is used to reproduce the turbulent viscosity model introduced in \refe{drr:eq:nut}, which will show that simple DL networks are able to learn from the \rebuttal{considered} DNS data. Afterwards, several regression models for the scalar dissipation rate are evaluated. All cases were run for 7000 epochs, and the evolution\rebuttal{s} of the loss functions are shown to evaluate the convergence of the training.

\subsubsection{$\nu_\mathrm{T}$ prediction using ${\bar{\tens{S}}}$:}
\begin{figure}[htbp]
	\centering 
	\includegraphics[scale=0.15]{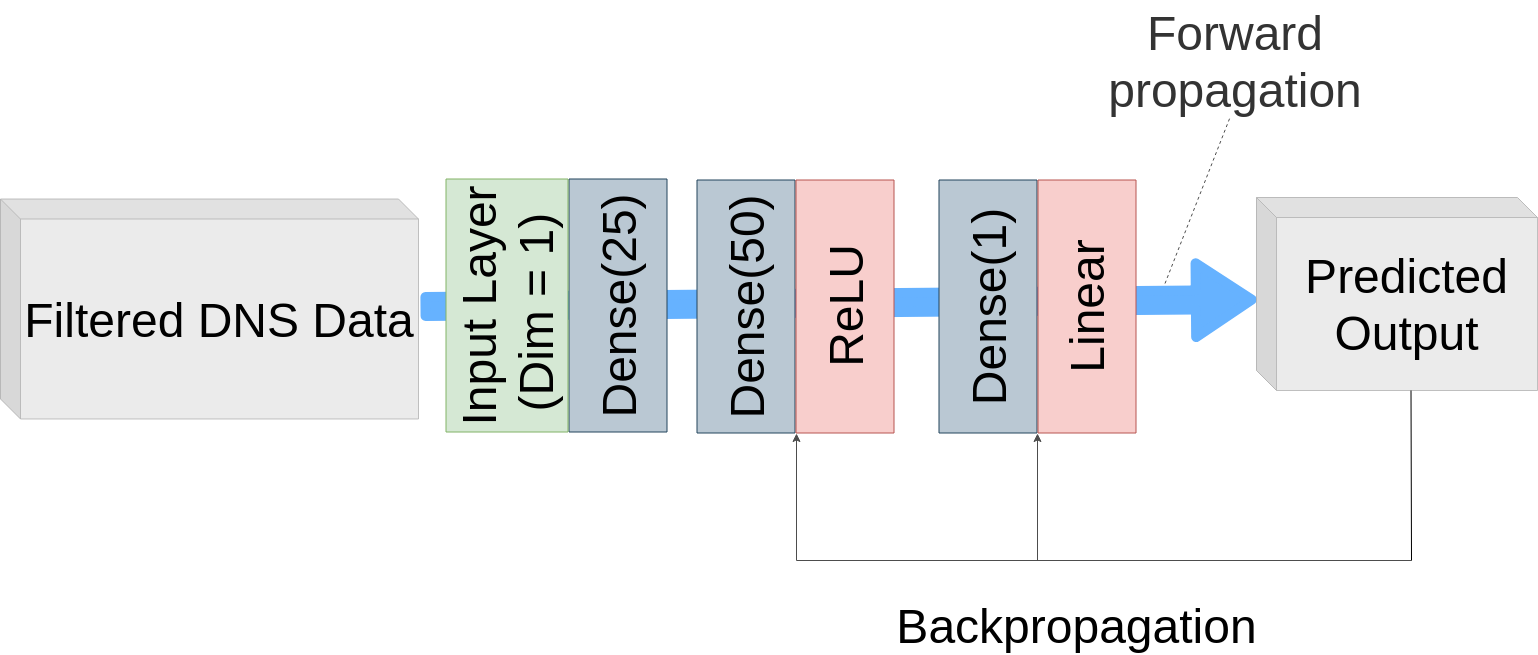}
	\caption{Sketch of the network for $\nu_\mathrm{T}$ prediction.}
	\label{drr:fig:network1}
\end{figure}

\begin{figure}[htbp]
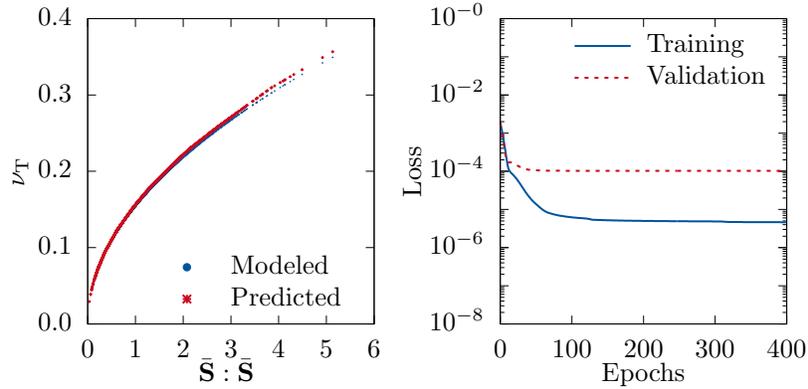

\picsize
	\centering
	\begin{subfigure}[t]{0.44\linewidth}
          \picbox{\input{figures/drr_pred_nut_tex.tex}}
	\end{subfigure}
	\begin{subfigure}[t]{0.44\linewidth}
		\picbox{\input{figures/drr_train_nut_tex.tex}}
	\end{subfigure}
	\caption{Visualization of the modeled and DL-predicted turbulent viscosity $\nu_\mathrm{T}$ using the double inner product of two filtered rate of strain tensors  $\bar{\tens{S}}:\bar{\tens{S}}$ as input (left) and the corresponding loss as function of number of epochs (right).}
	\label{drr:fig:nut}
\end{figure}

As network validation, a single input value, single output value mapping was implemented relating $\bar{\tens{S}}:\bar{\tens{S}}$ and $\nu_\mathrm{T}$ by means of a 3-layer neural network as shown in \reff{drr:fig:network1}. \reff{drr:fig:nut} compares the modeled $\nu_T$, obtained from
\refe{drr:eq:nut}, with the prediction from the network. The good collapse of both
curves for all values of $\bar{\tens{S}}:\bar{\tens{S}}$ validates that the network is able to learn simple relations as given by \refe{drr:eq:nut}.

%

\subsubsection{$\chi$ prediction using $\bar{\phi}$:}
\begin{figure}[htbp]
	\centering 
	\includegraphics[scale=0.15]{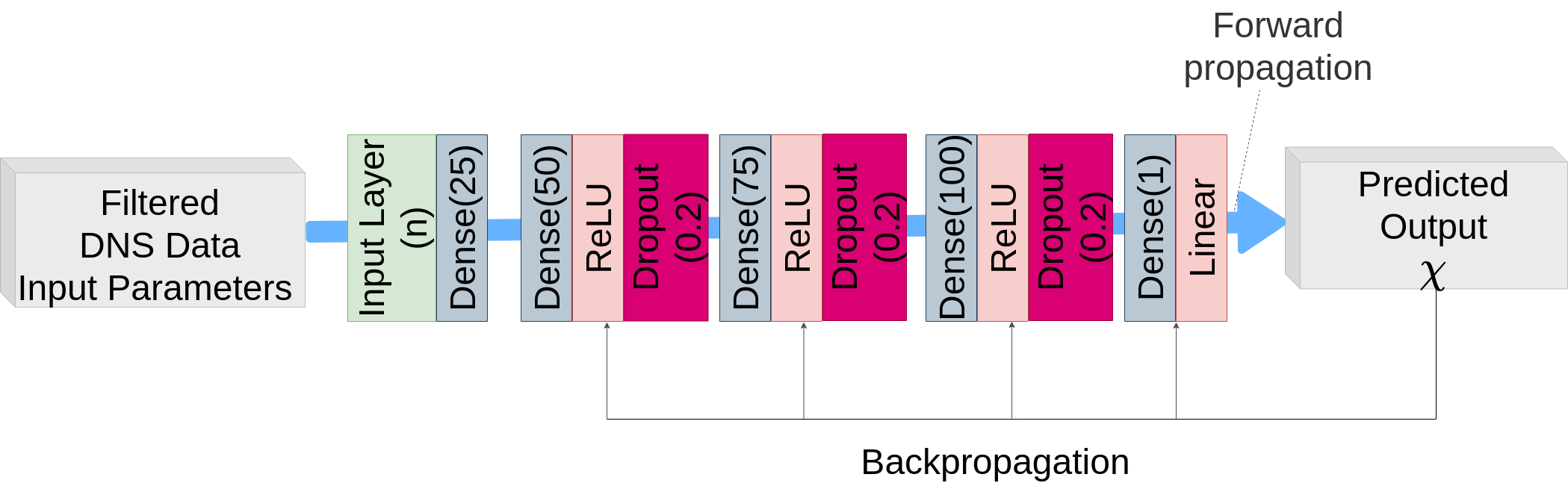}
	\caption{Sketch of the network for $\chi$ prediction.}
	\label{drr:fig:network2}
\end{figure}

After validating the network with predicting $\nu_\mathrm{T}$, feedforward networks are used to predict the resolved scalar dissipation rate $\chi$. The accuracy of the prediction is strongly affected by the \rebuttal{considered} input variables and the network architecture and parameters. It was found that the 3-layer network shown in \reff{drr:fig:network1}, which works well for predicting $\nu_\mathrm{T}$, leads to inaccuracies for predictions of the resolved scalar dissipation rate. The accuracy could be improved by switching to a 5-layer network architecture as visualized in \reff{drr:fig:network2}. Even more layers did not improve the prediction accuracy further, and therefore, the following plots are based on training with the 5-layer network.

\begin{figure}[H]
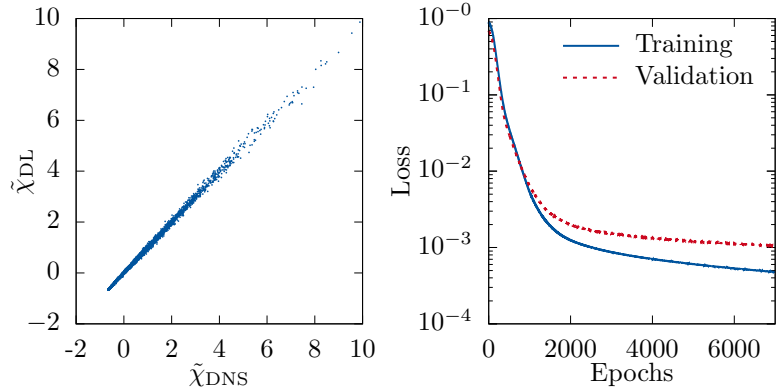

\picsize
	\centering
	\begin{subfigure}[t]{0.44\linewidth}
		\centering 
		\picbox{\input{figures/drr_pred_chi_ps_tex.tex}}
		\label{drr:fig:reg_chi_ps_result}
	\end{subfigure}
	\begin{subfigure}[t]{0.44\linewidth}
		\centering 
		\picbox{\input{figures/drr_train_chi_ps_tex.tex}}
		\label{drr:fig:reg_chi_ps_train}
	\end{subfigure}
	\caption{Visualization of the correlation between DNS and DL-predicted rescaled scalar dissipation rate \rebuttal{$\tilde{\chi}$} using the filtered passive scalar $\bar{\phi}$ as input (left) and the corresponding loss as function of number of epochs (right).}
	\label{drr:fig:reg_chi_ps}
\end{figure}

\begin{figure}[htbp]
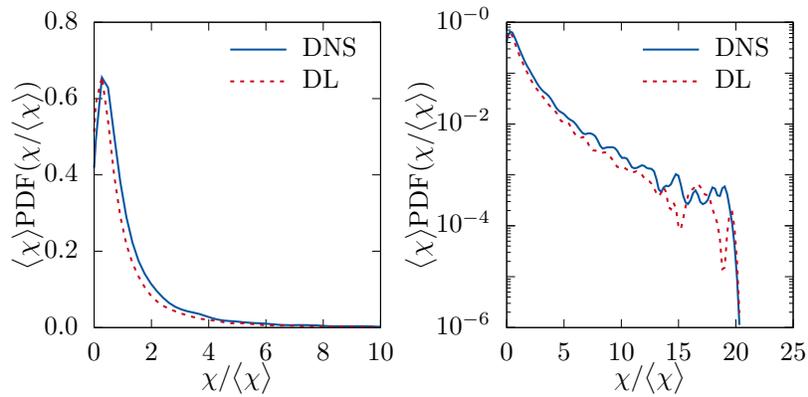

\picsize
	\centering	
	\begin{subfigure}[t]{0.44\textwidth}
		\centering 
		\picbox{\input{figures/drr_pdf_ps_tex.tex}}
		\label{drr:fig:reg_chi_ps_pdf_1}
	\end{subfigure}
	\begin{subfigure}[t]{0.44\textwidth}
		\centering 
		\picbox{\input{figures/drr_pdf_log_ps_tex.tex}}
		\label{drr:fig:reg_chi_ps_pdf_2}
	\end{subfigure}
	\caption{Visualization of the normalized PDF of the DNS and DL-predicted scalar dissipation rate $\chi$ with linear (left) and logarithmic (right) ordinate for the network with filtered passive scalar $\bar{\phi}$ as input.}
	\label{drr:fig:reg_chi_ps_pdf}
\end{figure}

The simplest approach is to use the filtered scalar $\bar{\phi}$ as only input to the network\rebuttal{, which is equal to the filtered scalar fluctuations $\overline{\phi-\big< \phi \big>}$ here}. The obtained results are shown in
\reff{drr:fig:reg_chi_ps}, and good correlation between the DNS and the DL-predicted values of $\chi$ can be seen. \rebuttal{Note that the negative values of the scalar dissipation rate result from a centering and rescaling of the scalar dissipation rate fields indicated by the tilde symbol.} \rebuttal{The good correlation} implies that the network is able to learn the derivatives of \rebuttal{$(\phi-\big< \phi \big>)$} (cf. \refe{drr:eq:sd_1}), even though no convolutional layer was used here. Moreover, the probability density function (PDF) of $\chi$ is plotted in \reff{drr:fig:reg_chi_ps_pdf} to further assess the
accuracy of the prediction. The
scalar dissipation rate is a very intermittent quantity, which implies the
presence of very strong but very rare
events.  These strong events are characteristic features of
turbulence and play an important role for small-scale mixing
or turbulent combustion. Comparing the PDFs of the DNS and
DL-predicted scalar dissipation rates indicates that the dense fully connected
neural network is able to reproduce the PDF of $\chi$ with moderate accuracy as clear deviations are seen in the logarithmic plot.

\subsubsection{$\chi$ prediction using $\bar{\phi}$ and $\bar{\varepsilon}$:}
Classical models in turbulence propose that the mean scalar
dissipation rate $\big< \chi \big>$ depends on the scalar variance
\rebuttal{$\big< (\phi-\big< \phi \big>)^2 \big>$} and a characteristic time-scale $\tau$, i.e.
\begin{equation}
  \label{drr:eq:sd2}
  \big< \chi \big> = c_\chi \frac{\big< (\phi-\big< \phi \big>)^2 \big>}{\tau} \,,
\end{equation}
where $c_\chi$ is a constant. \press{$\tau$ is usually chosen as an integral time-scale and can be defined as
\begin{equation}
\tau = \frac{\big< {k} \big>}{\big< {\varepsilon} \big>}.
\end{equation}
The integral
time-scale is a characteristic time-scale of the larger
eddies in a turbulent flow, which determine the rate of turbulent mixing.}

\begin{figure}[H]
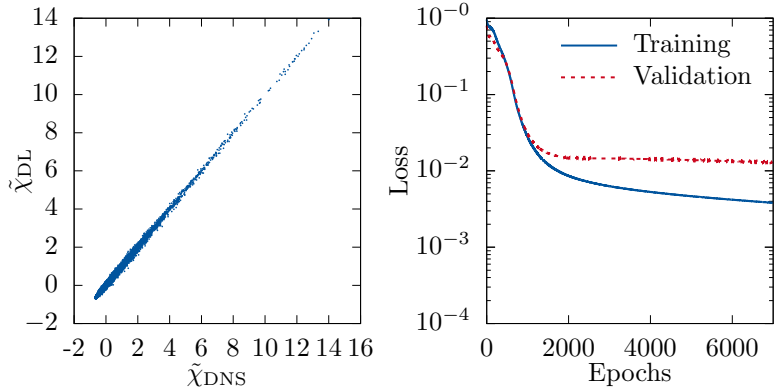

\picsize
	\centering
	\begin{subfigure}[t]{0.44\linewidth}
		\centering 
		\picbox{\input{figures/drr_pred_chi_ps_ed_tex.tex}}
		\label{drr:fig:reg_chi_ps_ed_result}
	\end{subfigure}
	\begin{subfigure}[t]{0.44\linewidth}
		\centering 
		\picbox{\input{figures/drr_train_chi_ps_ed_tex.tex}}
		\label{drr:fig:reg_chi_ps_ed_train}
	\end{subfigure}
	\caption{Visualization of the correlation between DNS and DL-predicted rescaled scalar dissipation rate \rebuttal{$\tilde{\chi}$} using the filtered passive scalar $\bar{\phi}$ and filtered energy dissipation rate $\bar{\varepsilon}$ as inputs (left) and the corresponding loss as function of number of epochs (right).}
	\label{drr:fig:reg_chi_ps_ed}
\end{figure}

\begin{figure}[htbp]
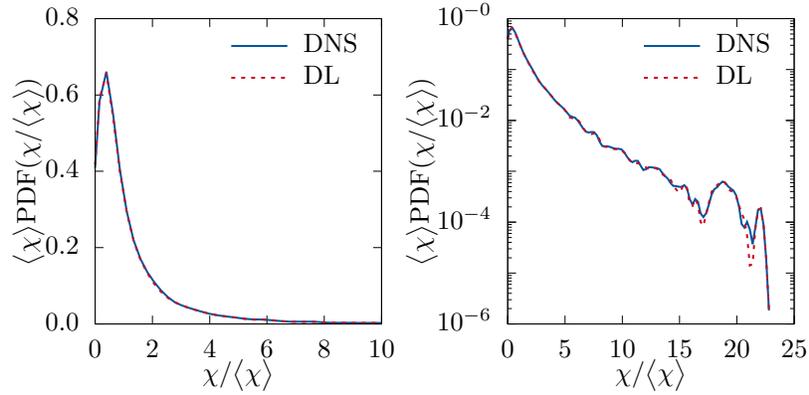

\picsize
	\centering
	\begin{subfigure}[t]{0.44\textwidth}
		\centering 
		\picbox{\input{figures/drr_pdf_ps_ed_tex.tex}}
		\label{drr:fig:reg_chi_ps_ed_pdf_1}
	\end{subfigure}
	\begin{subfigure}[t]{0.44\textwidth}
		\centering 
		\picbox{\input{figures/drr_pdf_log_ps_ed_tex.tex}}
		\label{drr:fig:reg_chi_ps_ed_pdf_2}
	\end{subfigure}
	\caption{Visualization of the normalized PDF of the DNS and DL-predicted scalar dissipation rate $\chi$ with linear (left) and logarithmic (right) ordinate for the network with filtered passive scalar $\bar{\phi}$ and filtered energy dissipation rate $\bar{\varepsilon}$ as inputs.}
	\label{drr:fig:reg_chi_ps_ed_pdf}
\end{figure}

Inspecting the relation given by \refe{drr:eq:sd2} insinuates that
the mapping shown in \reff{drr:fig:reg_chi_ps} and \reff{drr:fig:reg_chi_ps_pdf} may be
incomplete, since it neglects the dependence of $\chi$ on the
characteristic time-scale $\tau$, which leads to the observed deviations. \press{Following Overholt and Pope~\cite{Overholt1996} and motivated by \refe{drr:eq:sd2}}, the input for the network predicting the scalar dissipation rate $\chi$
is extended by the \rebuttal{resolved} energy dissipation rate $\bar{\varepsilon}$, defined as
\rebuttal{\begin{equation}
   \bar{\varepsilon} = \frac{1}{2} \nu \left(
    \nabla( \overline{\vect{u}-\big< \vect{u} \big>}) + (\nabla( \overline{\vect{u}-\big< \vect{u} \big>}))^\intercal \right):\left(
    \nabla( \overline{\vect{u}-\big< \vect{u} \big>}) + (\nabla( \overline{\vect{u}-\big< \vect{u} \big>}))^\intercal \right),
  \label{drr:eq:vareps_formal}
\end{equation}
which simplifies to
\begin{equation}
  \bar{\varepsilon} = 2 \nu \bar{\tens{S}}:\bar{\tens{S}}
  \label{drr:eq:vareps}
\end{equation}
for the data considered in this work.}
As can be seen in \reff{drr:fig:reg_chi_ps_ed} and \reff{drr:fig:reg_chi_ps_ed_pdf}, the prediction quality improves, probably because $\bar{\varepsilon}$ provides additional information about the local time scales of turbulence to the network.

\subsubsection{$\chi$ prediction using $\bar{\phi}$ and \rebuttal{$\bar{\vect{u}}$:}}
After successfully predicting the scalar dissipation rate with good accuracy, it is tested whether the network is also able to extract the time scale information contained in the filtered energy dissipation rate from the filtered velocity $\bar{\vect{u}}$, which is used to compute the filtered energy dissipation rate (cf. \refe{drr:eq:S} and \refe{drr:eq:vareps}). Therefore, the network inputs are changed to $\bar{\phi}$ and $\bar{\vect{u}}$, and the results are shown in \reff{drr:fig:reg_chi_ps_ui} and \reff{drr:fig:reg_chi_ps_ui_pdf}. It can be seen that the prediction quality is worse compared to \reff{drr:fig:reg_chi_ps_ed} and \reff{drr:fig:reg_chi_ps_ed_pdf}, which indicates that the network is not fully able to learn the tensor operations performed in \refe{drr:eq:S} and \refe{drr:eq:vareps}. \rebuttal{Interestingly, the result is also worse than the results shown in \reff{drr:fig:reg_chi_ps} and \reff{drr:fig:reg_chi_ps_pdf}, which might be due to overfitting.}
\begin{figure}[H]
\picsize
	\centering
	\begin{subfigure}[t]{0.44\linewidth}
		\centering 
		\picbox{\input{figures/drr_pred_chi_ps_ui_tex.tex}}
		\label{drr:fig:reg_chi_ps_ui_result}
	\end{subfigure}
	\begin{subfigure}[t]{0.44\linewidth}
		\centering 
		\picbox{\input{figures/drr_train_chi_ps_ui_tex.tex}}
		\label{drr:fig:reg_chi_ps_ui_train}
	\end{subfigure}
	\caption{Visualization of the correlation between DNS and DL-predicted rescaled scalar dissipation rate \rebuttal{$\tilde{\chi}$} using the filtered passive scalar $\bar{\phi}$ and filtered velocity $\bar{\vect{u}}$ as inputs (left) and the corresponding loss as function of number of epochs (right).}
	\label{drr:fig:reg_chi_ps_ui}
\end{figure}
\begin{figure}[htbp]
\picsize
	\centering
	\begin{subfigure}[t]{0.44\textwidth}
		\centering 
		\picbox{\input{figures/drr_pdf_ps_ui_tex.tex}}
		\label{drr:fig:reg_chi_ps_ui_pdf_1}
	\end{subfigure}
	\begin{subfigure}[t]{0.44\textwidth}
		\centering 
		\picbox{\input{figures/drr_pdf_log_ps_ui_tex.tex}}
		\label{drr:fig:reg_chi_ps_ui_pdf_2}
	\end{subfigure}
	\caption{Visualization of the normalized PDF of the DNS and DL-predicted scalar dissipation rate $\chi$ with linear (left) and logarithmic (right) ordinate for the network with filtered passive scalar $\bar{\phi}$ and filtered velocity $\bar{\vect{u}}$ as inputs.}
	\label{drr:fig:reg_chi_ps_ui_pdf}
\end{figure}

\subsection{Reconstruction}
%
Reconstructing the fully-resolved flow from large-scale or
coarse-grained data has significant applications in various
domains. For example, particle image velocimetry (PIV) measurements
can only resolve information on large scales due to limited spatial
resolution~\cite{cao2000piv}. Similarly, LES is widely used for
weather predictions~\cite{rotunno2009large}, where resolving the
small-scale information is prohibitively expensive. The reconstruction of subgrid information
with deep learning networks is a promising approach to link the large-scale results obtained from experiments or filtered equations to the actual flow fields.

In this subsection, a GAN-approach is used to reconstruct fully-resolved 3-D velocity fields from filtered data. With these fields, the filtered NSEs can be closed.

\subsubsection{Network motivation:}
The DL network used for reconstruction in this work is inspired by the enhanced
super-resolution GAN
({ESRGAN}) introduced by Wang et al.~\cite{wang2019} for reconstructing filtered features in 2-D images, which is a leading DL-approach in the field of single image super-resolution (SISR). A pioneering work in the field of SISR was
the SRCNN proposed by Dong et al.~\cite{dong2014learning}. The general concept of GANs was
presented by Goodfellow et al.~\cite{goodfellow2014generative}. A GAN
is composed of two models, a generator that captures the data
distribution and generates new data, and a discriminator that learns
to distinguish whether a sample stems from the original data
distribution (genuine) or the generator (fake). During training, the
generator learns to produce samples that are indistinguishable for the
discriminator, while the discriminator learns to more accurately judge
the genuineness.
For better perceptual
similarities, Ledig et al.~\cite{ledig2017photo} introduced the SRGAN,
which takes the perceptual loss into consideration while evaluating
the cost function. Instead of calculating the root-mean-square error (RMSE) in pixel space,
the content loss is implied by calculating the RMSE in
VGG19~\cite{simonyan2014very} feature space, i.\,e. the VGG loss. This
grants the SR-images produced by a SRGAN generator satisfying perceptual
similarity to the original image as well as optimized recovery of the
high frequency details. However, SRGAN produced hallucinated details
accompanied with unpleasant artifacts in the
images~\cite{wang2019}. Hence, Wang et al. proposed the Enhanced
SRGAN (ESRGAN) to alleviate such problems by building a
residual-in-residual dense block (RRDB) into the SRGAN generator and
adopting the idea of relativistic
GAN~\cite{jolicoeur2018relativistic}.

\begin{figure}[htbp]
 \centering
\includegraphics[scale=0.15]{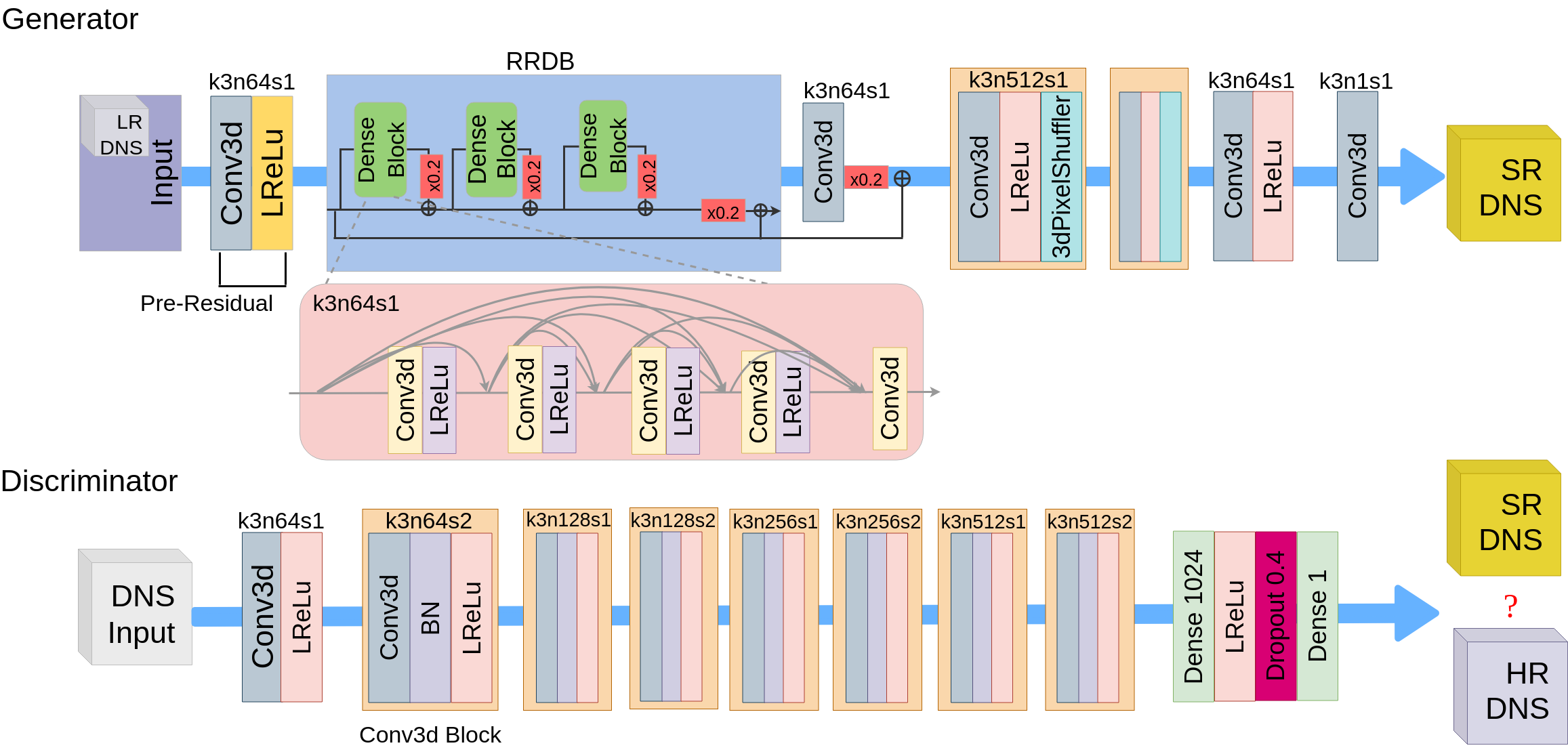}
 \caption{Sketch of the network used for the reconstruction.}
 \label{drr:fig:rec}
\end{figure}

\begin{figure}[htbp]
\picsize
	\centering
	\begin{subfigure}[t]{0.30\linewidth}
		\centering 
		\picbox{\includegraphics[width=\linewidth]{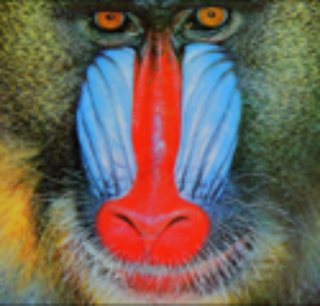}}
		\label{drr:fig:tsrgan_val_org}
	\end{subfigure}
	\hspace{1mm}
	\begin{subfigure}[t]{0.30\linewidth}
		\centering 
		\picbox{\includegraphics[width=\linewidth]{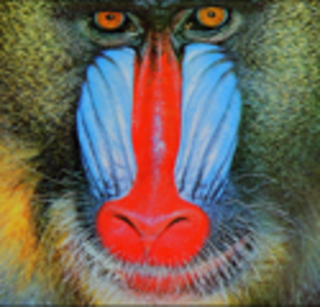}}
		\label{drr:fig:tsrgan_val_bi}
	\end{subfigure}
	\hspace{1mm}
	\begin{subfigure}[t]{0.30\linewidth}
		\centering 
		\picbox{\includegraphics[width=\linewidth]{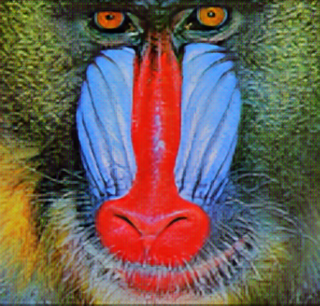}}
		\label{drr:fig:tsrgan_val_gan}
	\end{subfigure}
	\caption{Comparison of an original (left), bicubic interpolated (center), and TSRGAN-reconstructed image. \rebuttal{The original image is taken from the DIV2K archive~\cite{Agustsson_2017_CVPR_Workshops}.}}
	\label{drr:fig:esrgan_test}
\end{figure}

The ESRGAN has been extended to a turbulence super-resolution
GAN (TSRGAN) for this work, as shown in \reff{drr:fig:rec}. The TSRGAN is able to deal with 3-D subboxes of the filtered DNS data (scalar and vector fields) as input and employs physics-based loss functions for
training of the network. Validation results of the TSRGAN trained with
\num{800} images from the DIV2K archive~\cite{Agustsson_2017_CVPR_Workshops} over \num{50000} epochs are presented in \reff{drr:fig:esrgan_test}. Besides the good quality of 2-D reconstruction on images, also the similarity in terms of tensor operations seems to make the TSRGAN a promising candidate for reconstruction of filtered flow data. A filter operation can be seen as convolution, and the network architecture of the TSRGAN heavily relies on convolutional layers.
%

\subsubsection{Loss function:}
The perceptual loss proposed for the ESRGAN based on VGG-feature space
is apparently not as suitable for the turbulence data, as the
geometrical features from VGG19 are not representative for
turbulent flows. Hence, a new formulation for the cost
function was developed inspired by physical flow constraints.

Before training the TSRGAN as a combined model, the generator is
pretrained with RMSE due to the complexity of the RRDB.
For the combined model, the loss function for reconstructing velocity fields is proposed as 
\begin{equation}
l = \beta_1 l_\mathrm{RADG} + \beta_2 l_\mathrm{pixel} + \beta_3 l_\mathrm{gradient} + \beta_4 l_\mathrm{continuity}
\end{equation}
with $\beta_1, \beta_2, \beta_3$, and $\beta_4$ being coefficients weighting the different loss term contributions. $l_\mathrm{RADG}$ is the 'realistic average' discriminator/generator loss, which is the accuracy feedback between discriminator and generator as given by Wang et al.~\cite{wang2019}. The pixel loss $l_\mathrm{pixel}$ is defined as
\begin{equation}
l_\mathrm{pixel} = \mathrm{MSE}(\vect{u}^{\mathrm{predicted}},\vect{u}^{\mathrm{DNS}}).
\end{equation}
The mean-scare error (MSE) operator is given by
\begin{equation}
\mathrm{MSE}(\{\cdot\}_1,\{\cdot\}_2)=\frac{1}{N_{\mathrm{samples}}} \sum_{i=1}^{N_\mathrm{samples}}(\{\cdot\}_1^i - \{\cdot\}_2^i)^2
\end{equation}
with $N_{\mathrm{samples}}$ as number of all samples, i.\,e. the total number of grid points of the reconstructed field. If the MSE operator is applied on tensors including vectors, it is applied to all elements separately. Afterwards the resulting tensor is mapped into a scalar using the 1-norm. The gradient loss $l_\mathrm{gradient}$ is defined as
\begin{equation}
l_\mathrm{gradient} = \mathrm{MSE}(\nabla\vect{u}^{\mathrm{predicted}},\nabla\vect{u}^{\mathrm{DNS}}).
\end{equation}
$l_\mathrm{continuity}$ is the continuity loss, which enforces the continuity equation in the reconstructed field and reads
\begin{equation}
l_\mathrm{continuity} = \mathrm{MSE}(\nabla\cdot\vect{u}^{\mathrm{predicted}},\vect{0}).
\end{equation}

%

\subsubsection{Results:}
\begin{figure}[htbp]
\picsize
	\centering
	\begin{subfigure}[t]{0.30\linewidth}
		\centering 
		\picbox{\includegraphics[trim=47 15 52 25, clip,width=\linewidth]{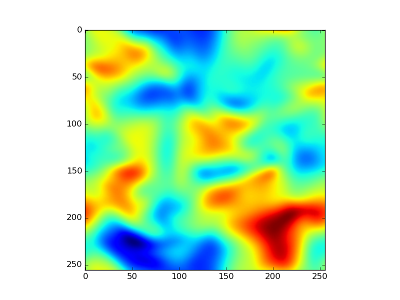}}
		\label{drr:fig:filt_u}
	\end{subfigure}
	\hspace{1mm}
	\begin{subfigure}[t]{0.30\linewidth}
		\centering 
		\picbox{\includegraphics[trim=47 15 52 25, clip,width=\linewidth]{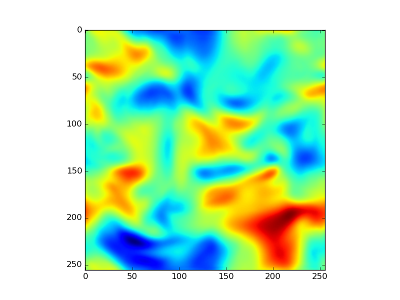}}
		\label{drr:fig:dl_u}
	\end{subfigure}
	\hspace{1mm}
	\begin{subfigure}[t]{0.30\linewidth}
		\centering 
		\picbox{\includegraphics[trim=47 15 52 25, clip,width=\linewidth]{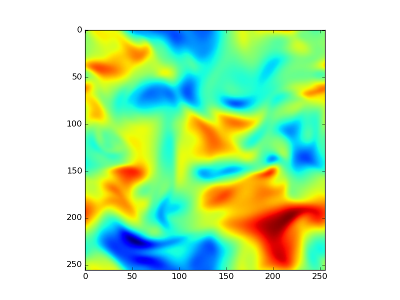}}
		\label{drr:fig:dns_u}
	\end{subfigure}
	
	\begin{subfigure}[t]{0.30\linewidth}
		\centering 
		\picbox{\includegraphics[trim=47 15 52 25, clip,width=\linewidth]{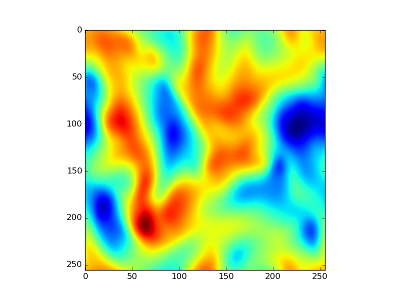}}
		\label{drr:fig:filt_v}
	\end{subfigure}
	\hspace{1mm}
	\begin{subfigure}[t]{0.30\linewidth}
		\centering 
		\picbox{\includegraphics[trim=47 15 52 25, clip,width=\linewidth]{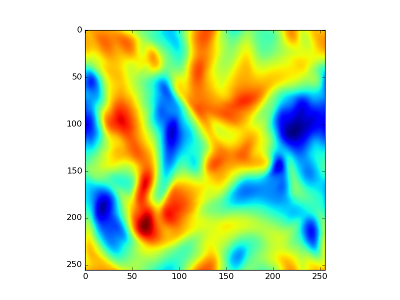}}
		\label{drr:fig:dl_v}
	\end{subfigure}
	\hspace{1mm}
	\begin{subfigure}[t]{0.30\linewidth}
		\centering 
		\picbox{\includegraphics[trim=47 15 52 25, clip,width=\linewidth]{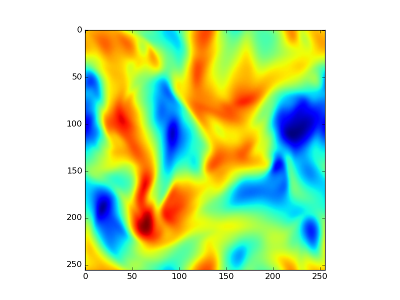}}
		\label{drr:fig:dns_v}
	\end{subfigure}
	
    \begin{subfigure}[t]{0.30\linewidth}
		\centering 
		\picbox{\includegraphics[trim=47 15 52 25, clip,width=\linewidth]{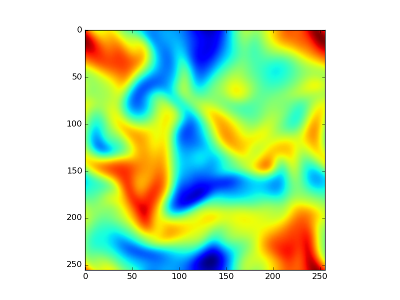}}
		\label{drr:fig:filt_w}
	\end{subfigure}
	\hspace{1mm}
	\begin{subfigure}[t]{0.30\linewidth}
		\centering 
		\picbox{\includegraphics[trim=47 15 52 25, clip,width=\linewidth]{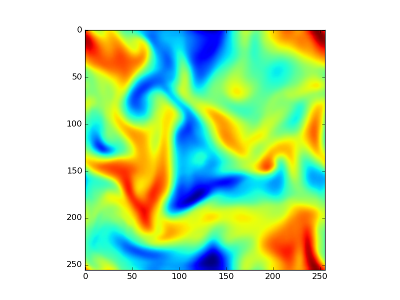}}
		\label{drr:fig:dl_w}
	\end{subfigure}
	\hspace{1mm}
	\begin{subfigure}[t]{0.30\linewidth}
		\centering 
		\picbox{\includegraphics[trim=47 15 52 25, clip,width=\linewidth]{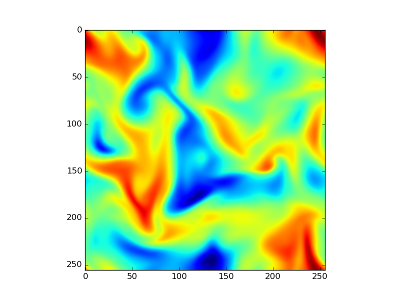}}
		\label{drr:fig:dns_w}
	\end{subfigure}
	\caption{Comparison of 2-D slices of the filtered (left), DL-reconstructed (center),  and DNS (right) data. Snapshots of the three elements of the velocity vector $\vect{u}$ are shown row-by-row.}
	\label{drr:fig:vis_u}
\end{figure}

\begin{figure}[htbp]
\picsize
	\centering
    \begin{subfigure}[t]{0.30\linewidth}
		\centering 
		\picbox{\includegraphics[trim=47 15 52 25, clip,width=\linewidth]{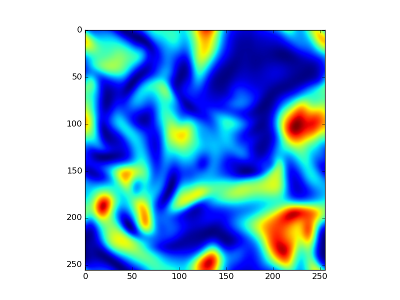}}
		\label{drr:fig:filt_k}
	\end{subfigure}
	\hspace{1mm}
	\begin{subfigure}[t]{0.30\linewidth}
		\centering 
		\picbox{\includegraphics[trim=47 15 52 25, clip,width=\linewidth]{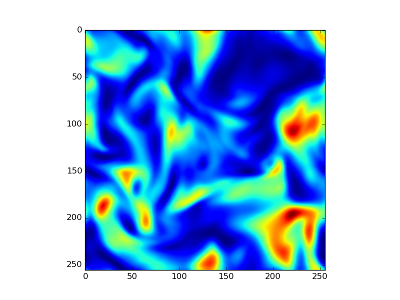}}
		\label{drr:fig:dl_k}
	\end{subfigure}
	\hspace{1mm}
	\begin{subfigure}[t]{0.30\linewidth}
		\centering 
		\picbox{\includegraphics[trim=47 15 52 25, clip,width=\linewidth]{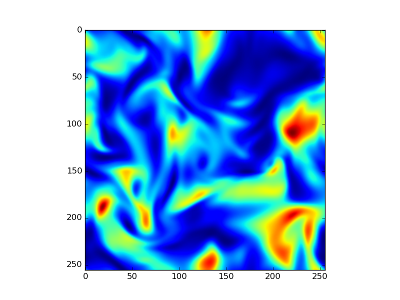}}
		\label{drr:fig:dns_k}
	\end{subfigure}
	\caption{Comparison of 2-D slices of turbulent \rebuttal{kinetic} energy $k$ snapshots for filtered (left), DL-reconstructed (center),  and DNS (right) data.}
	\label{drr:fig:vis_k}
\end{figure}

\begin{figure}[htbp]
\picsize
	\centering
	\begin{subfigure}[t]{0.44\linewidth}
		\centering 
		\picbox{\includegraphics[width=\linewidth]{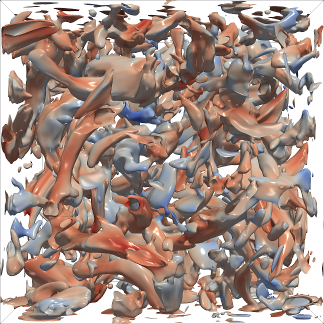}}
		\label{drr:fig:reg_q_dns}
	\end{subfigure}
	\hspace{1mm}
	\begin{subfigure}[t]{0.44\linewidth}
		\centering 
		\picbox{\includegraphics[width=\linewidth]{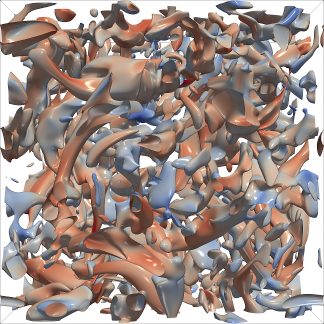}}
		\label{drr:fig:reg_q_dl}
	\end{subfigure}
	\caption{Comparison of the Q-criterion evaluated on the DL-reconstructed (left) and DNS (right) data.}
	\label{drr:fig:Q}
\end{figure}

To assess the performance of the TSRGAN, the network is trained with Case A and evaluated on Case B. \reff{drr:fig:vis_u} shows 2-D slices of the original DNS velocity fields, the filtered velocity fields, and reconstructed velocity fields. Additionally, \reff{drr:fig:vis_k} shows 2-D slices of turbulent \rebuttal{kinetic} energy snapshots. It is clearly visible that small-scale structures are missing in the filtered data. The TSRGAN predicts these
structures based on the large-scale features that are present in the
filtered field, and the visual agreement between DNS and the predicted
solution is very good. Moreover, \reff{drr:fig:Q} shows the
vortex structure of the DNS and reconstructed velocity fields, defined by the Q-criterion~\cite{dubief2000coherent}. The Q-criterion identifies coherent vortex structures by an
iso-surface of
\begin{equation}
  Q= \frac{1}{4} \left(\vecg{\textomega}\cdot\vecg{\textomega} -2 \tens{S}:\tens{S} \right),
\end{equation}
where $\vecg{\textomega}\cdot\vecg{\textomega}$ is the enstrophy. By definition, $Q$ is a small-scale
quantity, which is suitable to assess the turbulent motions in the
dissipative range. The agreement between DNS and reconstructed data is good.

\reff{drr:fig:vis_u}, \reff{drr:fig:vis_k}, and \reff{drr:fig:Q} show good visual agreement between DNS and reconstructed data. However, as turbulence is \rebuttal{a} multi-scale phenomenon, a visual evaluation of turbulence fields is often misleading, and a statistical assessment is necessary. The spectrum of the turbulent \rebuttal{kinetic} energy $E(\kappa)$ is
a statistical representation of the turbulent \rebuttal{kinetic} energy in wavenumber
space. Different scales can be distinguished: the energy-containing
range at small wavenumbers, the inertial subrange at intermediate
wavenumbers, and the dissipative range at large wavenumbers. However,
it is important to emphasize that a well defined scale-separation
between small and large scales only exists at sufficiently high
Reynolds numbers. When $E(\kappa)$ is known, the mean turbulent
energy can be obtained by
\begin{equation}
  \label{eq:Ek}
  \big< k \big> = \int_0^\infty E(\kappa) \di \kappa,
\end{equation}
whereas the mean energy dissipation rate equals
\begin{equation}
  \label{eq:Eeps}
  \big< \varepsilon \big> = 2\nu \int_0^\infty \kappa^2 E(\kappa) \di \kappa.
\end{equation}
In the context of LES, the filtering operation acts like a low-pass
filter and mainly affects the dissipative range, which in turn has a
stronger impact on the filtered mean dissipation rate
$\big< \bar \varepsilon \big>$ than on the filtered mean turbulent
energy \rebuttal{$\big< \bar{k} \big>$}. 

\begin{figure}[htbp]
\picsize
 \centering
\picbox{\input{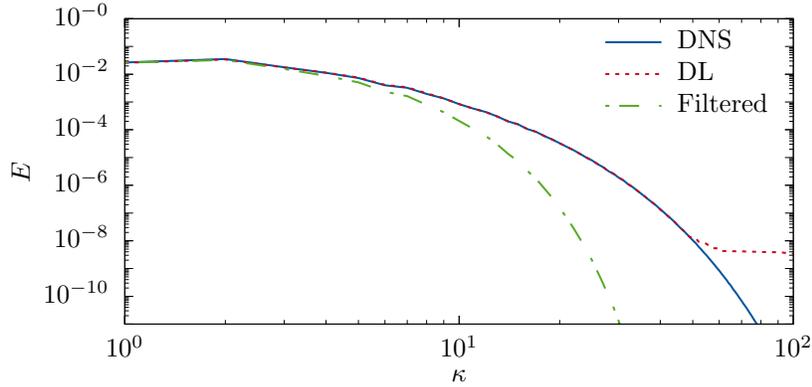}}
	\caption{Comparison of the energy spectra $E(\kappa)$ evaluated on the DNS, DL-reconstructed, and filtered data.}
	\label{drr:fig:reg_spec}
\end{figure}

\reff{drr:fig:reg_spec} compares the energy spectrum evaluated on the DNS, reconstructed, and filtered data. It can be
observed that the filtering operation is limited to the large
wavenumbers and \rebuttal{that it} removes most energy from the dissipative
range. The TSRGAN is able to predict these scales resulting in good agreement between DL-predicted and DNS spectra, except for very large wave numbers in the far dissipative
range, where the TSRGAN slightly over-predicts the turbulent
energy. These findings support the hypothesis that the TSRGAN is able to learn and reproduce features of small-scale turbulence and can be used to close the LES equations.

\section{Computing}
\label{drr:sec:C}
Typically, the single node performance of DL training is very good due to the heavy use of linear algebra-based primitives and the optimization of current GPUs for tensor operations. This is especially true if state-of-the-art libraries, such as TensorFlow, which are highly optimized for GPU usage, are used, as in this work. However, HPC-DL is still challenging. A common way for parallelizing the training of DL networks is to replicate the network across ranks. Thus, each rank processes a different local batch of DNS data, and updates to the network are aggregated among ranks during each training step.

For transforming single-process TensorFlow entities into a data-parallel implementation, Horovod~\cite{sergeev2018} was used, which adds allreduce operations into the back-propagation computation to average the computed gradients from each rank's network. The local networks are updated by the ranks independently, which results in synchronous distributed training due to the use of gradients averaged across all ranks. Obviously, two main challenges are the communication of the information and I/O of data for this procedure. They are addressed separately in the next two subsections. All highly-parallel training for this work was performed on the
Supercomputer JURECA at J\"ulich Supercomputing Centre (JSC), which features nodes equipped with two NVIDIA K80 GPUs (four visible devices per node). Finally, it was possible to train networks with up to \SI{396.2}{TFLOPS} on JURECA.

\subsection{Communication}
Horovod uses the first rank as central scheduler for all Horovod operations, employing a dynamical reordering of allreduce operations in order to achieve consistency among all ranks and avoid deadlock due to the independent scheduling of all TensorFlow entities. With an increasing number of ranks, the central scheduler becomes more and more a communication bottleneck as it needs to handle all readiness messages of all other ranks. As a distribution of this scheduling load is not possible due to the required total order of the collective operations, a communication tree was employed in this work. It allows to use Horovod's original scheduler but limits the message load due to the recursive broadcast.

\subsection{I/O}
As a large amount of DNS data is required for the training of the network, the data transfer to the GPUs is often a bottleneck as the file system - on JURECA GPFS is used - is not fast enough to feed the GPUs in a timely fashion. For this work, a similar strategy as suggested by Kurth et al.~\cite{kurth2018} was employed. Only a significant fraction of the overall data set was made accessible to each node for the distributed training setting. The locally available data were combined to a local batch in such a way that the set of samples for each rank was statistically similar to a batch selected from the entire data set. Technically,  a distributed data staging system was used that first divided the data set into disjoint pieces to be read by each rank, before distributing copies of each file to other nodes by point-to-point MPI messages. This approach takes advantage of the high bandwidth of the InfiniBand network without increasing the load on the file system.

\section{Conclusion}
Two DL approaches for modeling of subgrid statistics are presented in this paper. It is shown that simple feedforward ANNs are able to learn subgrid statistics with good accuracy if appropriate inputs are chosen. Furthermore, ESRGAN is extended to TSRGAN and used to reconstruct fully-resolved 3-D velocity fields. Both the visual agreement and the statistical agreement are very good, which indicates that the TSRGAN is able to predict small-scall turbulence. Finally, the code framework used for learning was optimized to achieve \SI{396.2}{TFLOPS} on the supercomputer JURECA.

\section*{Acknowledgment}
The authors gratefully acknowledge the computing time granted for the
project JHPC55 by the JARA-HPC Vergabegremium and provided on the
JARA-HPC Partition part of the supercomputer JURECA at
Forschungszentrum J\"ulich. Also financial support by the Cluster of Excellence “The Fuel Science Center”, which is funded by the Deutsche Forschungsgemeinschaft (DFG, German Research Foundation) under Germany’s Excellence Strategy – Exzellenzcluster 2186 “The Fuel Science Center” ID: 390919832, and from of the European
Research Council (ERC) under the European Union’s
Horizon 2020 research and innovation program under grant
agreement No 695747 is acknowledged. MG acknowledges financial support provided under the grant EMCO2RE. Furthermore, the authors want to thank Jenia Jitsev, Zeyu Lian, and Mayur Vikas Joshi for their help.

\bibliographystyle{splncs04}
\bibliography{bib/literature}
\end{document}